# Surface plasmon modes of a single silver nanorod: an electron energy loss study

Olivia Nicoletti,[1,*] Martijn Wubs,[2] N. Asger Mortensen,[2] Wilfried Sigle,[3] Peter A. van Aken,[3] and Paul A. Midgley[1]

[1]Department of Materials Science and Metallurgy, University of Cambridge, Pembroke Street, CB2 3QZ Cambridge, U.K.
[2]Department of Photonics Engineering, Technical University of Denmark, Ørsteds Plads 343, DK-2800 Kgs. Lyngby, Denmark
[3]Max Planck Institute for Intelligent Systems, Heisenbergstraße 3, D-70569 Stuttgart, Germany
[*]on214@cam.ac.uk

**Abstract:** We present an electron energy loss study using energy filtered TEM of spatially resolved surface plasmon excitations on a silver nanorod of aspect ratio 14.2 resting on a 30 nm thick silicon nitride membrane. Our results show that the excitation is quantized as resonant modes whose intensity maxima vary along the nanorod's length and whose wavelength becomes compressed towards the ends of the nanorod. Theoretical calculations modelling the surface plasmon response of the silver nanorod-silicon nitride system show the importance of including retardation and substrate effects in order to describe accurately the energy dispersion of the resonant modes.

## References and links

1. H. Raether, *Excitation of plasmons and Interband Transitions by Electrons* (Springer-Verlag, Berlin, 1980), Chap. 1.
2. H. Raether, *Surface Plasmons on Smooth and Rough Surfaces and on Gratings* (Springer-Verlag, Berlin, 1986), Chap. 2 and Chap. 3.
3. C. Nylander *et al*., "Gas detection by means of surface plasmon resonance." Sensors & Actuators **3**, 79-88, (1982).
4. D. Sarid and W. Channeler, *Surface Plasmons Theory, Mathematica Modeling, and Applications* (Cambridge University Press, Cambridge, 2010), Chap. 12.
5. W. L. Barnes and A. Dereux and T. Ebbesen, "Surface plasmon subwavelength optics." Nature **424,** 824-830 (2003).
6. F. J. García De Abajo, "Optical excitations in electron microscopy." Rev. Mod. Phys. **82,** 209-275 (2010).
7. P. C. Tiemeijer, "Operation modes of a TEM monochromator." Inst. Phys. Conf. Ser. **161,** 191–194, (1999).
8. R. F. Egerton, *Electron Energy-Loss Spectroscopy in the Electron Microscope – Second Edition* (Plenum Press, New York, 1996), Chap. 2.
9. J. Nelayah *et al.*, "Mapping surface plasmons on a single metallic nanoparticle." Nature Phys. **3,** 348-353, (2007).
10. J. Nelayah *et al.*, "Direct imaging of surface plasmon resonances on single triangular silver nanoprisms at optical wavelength using low-loss EFTEM imaging." Opt. Lett. **34,** 1003-1005, (2009).
11. W. Sigle and J. Nelayah and C. T. Koch and P. A. van Aken, "Electron energy losses in Ag nanoholes-from localized surface plasmon resonances to rings of fire." Opt. Lett. **34,** 2150-2152, (2009).
12. M. Bosman and V. J. Keast and M. Watanabe and A. I. Maaroof and M. B. Cortie, "Mapping surface plasmons at the nanometre scale with an electron beam." Nanotechnology **18,** 165505 (5pp) (2007).
13. B. Schaffer and U. Hohenester and A. Trügler and F. Hofer, "High-resolution surface plasmon imaging of gold nanoparticles by energy-filtered transmission electron microscopy." Phys. Rev. B **79,** 041401, (2009).
14. M. N'gom *et al.*, "Single Particle Plasmon Spectroscopy of Silver Nanowires and Gold Nanorods." Nano Lett. **8,** 3200-3204, (2008).
15. D. Rossouw and M. Couillard and J. Vickery and E. Kumacheva and G. A. Botton, "Multipolar Plasmonic Resonances in Silver Nanowire Antennas Imaged with a Subnanometer Electron Probe." Nano Lett. **11,** 1499–1504, (2011).

16. A. K. Koh *et al.*, "Electron Energy-Loss Spectroscopy (EELS) of Surface Plasmons in Single Silver Nanoparticles and Dimers: Influence of Beam Damage and Mapping of Dark Modes." ACS Nano **3,** 3015-3022, (2009).
17. M. –W. Chu *et al.*, "Probing Bright and Dark Surface-Plasmon Modes in Individual and Coupled Noble Metal Nanoparticles Using an Electron Beam." Nano Lett. **9,** 399-404 (2009).
18. D. B. Williams and C. B. Carter, *Transmission Electron Microscopy* (Springer, New York, 1996), Chap. 37 and Chap. 38.
19. C. T. Koch *et al.*, "SESAM: Exploring the Frontiers of Electron Microscopy." Microsc. Microanal. **12,** 506-514, (2006).
20. L. Novotny, "Effective Wavelength Scaling for Optical Nanoantennas." Phys. Rev. Lett. **98,** 266802 (4), (2007).
21. E. R. Encina and E. M. Perassi and E. A. Coronado, "Near-Field Enhancement of Multipole Plasmon Resonances in Ag and Au Nanowires." J. Phys. Chem. A **113,** 4489–4497 (2009).
22. N. Yamamoto *et al.*, "Light emission by surface plasmons on nanostructures of metal surfaces induced by high-energy electron beams." Surf. Interface Anal. **38,** 1725–1730, (2006).
23. E. J. R. Vesseur and R. de Waele and M. Kuttge and A. Polman, "Direct Observation of Plasmonic Modes in Au Nanowires Using High-Resolution Cathodoluminescence Spectroscopy." Nano Lett. **7,** 2843-2846, (2007).
24. R. Gómes-Medina and N. Yamamoto and M. Nakano and F. J. García de Abajo, "Mapping plasmons in nanoantennas via cathodoluminescence." New J. Phys. **10,** 105009, (2008).
25. K. E. Korte and S. E. Skrabalak Y. Xia, "Rapid synthesis of silver nanowires through a CuCl- or CuCl2-mediated polyol process." J. Mater. Chem. **18,** 437-441, (2008).
26. The silver nanowires used in this study were purchased from the Nano Research Facility (NRF), a member of the National Nanotechnology Infrastructure Network (NNIN), which is supported by the National Science Foundation under NSF award no. ECS-0335765. NRF is part of School of Engineering and Applied Science at Washington University in St. Louis.
27. E. Essers *et al.*, "Energy resolution of an Omega-type monochromator and imaging properties of the MANDOLINE filter." Ultramicroscopy **110,** 971-980, (2010).
28. B. Schaffer and W. Grogger and G. Kothleitner, "Automated spatial drift correction for EFTEM image series." Ultramicroscopy **102**, 27–36, (2004).
29. N. E. Christensen, "The Band Structure of Silver and Optical Interband Transitions." Phys. Stat. Sol. (b) **54,** 551-563 (1972).
30. B. E. Sernelius, *Surface Modes in Physics* (Wiley, New York, 2001), Chap. 7.
31. J. D. Jackson, *Classical Electrodynamics – Third Edition* (Wiley, New York, 1999), Chap. 6.
32. J. Grand *et al.*, "Role of localized surface plasmons in surface-enhanced Raman scattering of shape-controlled metallic particles in regular arrays." Phys. Rev. B **72,** 033407, (2005).
33. W. –B. Ewe and H. –S. Chu and E. –P. Li and B. S. Luk'yanchuk, "Field enhancement of gold optical nanoantennas mounted on a dielectric waveguide." Appl. Phys. A **100,** 315–319, (2010).
34. R. F. Egerton, "Limits to the spatial, energy and momentum resolution of electron energy-loss spectroscopy." Ultramicroscopy **107,** 575-586, (2007).

## 1. Introduction

Many of the remarkable optical properties offered by metallic nanoparticles (NPs) arise because of the excitation of surface plasmon (SP) resonances. Plasmons are the collective coherent excitation of conduction electrons [1] and surface plasmons are a sub-set whose nature is dictated by the interaction of conduction electrons with the interface between a metal and a dielectric medium [2]. The rich variety of SP resonant modes seen in metallic NPs is brought about by the dependence of SP excitation with the NP shape, size, composition and environment. Such dependence has led to many proposed applications, including (a) chemical and biochemical sensors which make use of the property that the energy at which the SPs occur depends on the dielectric function of the surrounding environment [3,4], (b) surface-enhanced Raman spectroscopy (SERS) substrates that rely on the ability of SPs to enhance the local electric field increasing enormously for example the Raman response of molecules, and (c) nanophotonic waveguides [4,5] which enables the SP to couple with surface roughness and be converted into electromagnetic radiation (and *vice versa*) [2,6].

Most of these applications are dependent on the sub-wavelength spatial variations of the SPs induced in the metal NPs. It is therefore of paramount importance to understand such variations using characterization techniques that offer sufficient spatial and energy resolution to access this information.

SPs have been studied for the most part by light optical excitation, using either reflective or absorption experiments [2]. Only in recent years, thanks primarily to improvements in energy resolution given by the introduction of electron monochromators on commercial transmission electron microscopes (TEMs) [7], has electron energy-loss spectroscopy (EELS) [8] begun to be used routinely as a complementary technique to light-induced SP excitation and analysis; recent publications on the direct mapping of SP resonant modes on metal nanoparticles by EELS include references [9-15]. The advantages of probing optical excitations with electrons include the possibility of much higher spatial resolution [6] (through the small De Broglie wavelength of the electron beam) and the ability to excite all possible SP modes (both bright and dark modes [16,17]).

All the measurements reported in this article are based on the technique of energy filtered TEM (EFTEM), where SP resonant modes are excited using parallel illumination with a series of images acquired, each of which is formed using electrons that have lost energies within a small range (in this study 0.23 eV), selected by an energy window in the spectral plane [8,18,19].

Metallic nanorods with "sub-wavelength" dimension (i.e. with dimensions smaller than the wavelength of emitted light) can be used as nano-antennae. Theoretical work [20,21] and initial experimental measurements [15,22-24] illustrate the great interest in, and potential of, such systems. We have therefore undertaken a detailed energy-loss spectroscopy study of an isolated silver nanorod acting as a nano-antenna, mapping the spatial variation of SP excitations along the nanorod using EFTEM and with detailed analysis of important spatial parameters and features of the SP resonance.

## 2. Materials and Methods

Silver nanoparticles in solution were synthesized via polyol synthesis, following the preparation reported in [25,26]. The majority of nanoparticles present in solution have an elongated rod shape, i.e. they are solid cylinders with approximately hemi-spherical ends, one example of which can be seen in Figure 1. The nanorods studied have aspect ratios varying from 2 to ca. 100, with the minor axis (diameter) of the rod being of the order of 50 nm for nearly all rods. A minority of particles have shapes of higher symmetry, such as triangular prisms, pyramids, cubes and spheres. For TEM sample preparation a drop of solution was dried on a 30 nm thick silicon nitride membrane purchased from Agar Scientific, used as a TEM substrate.

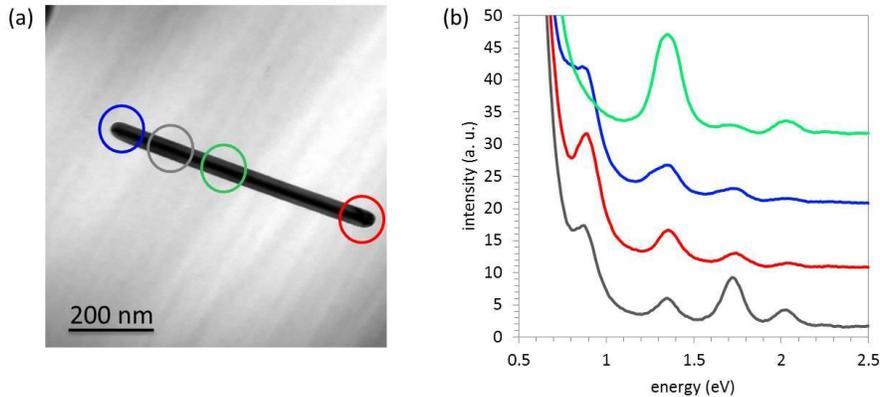

Figure 1. (a) Bright-field (BF) zero-loss image of a 666±3 nm long silver nanorod, with 47±3 nm diameter. (b) Electron energy-loss spectra (unprocessed) acquired at the positions marked in (a) of matching colour. Spectra have been obtained using a selected-area approach [18] and shown displaced on the intensity axis for clarity.

In EFTEM, a parallel electron beam interacts with the specimen and the transmitted electron beam is sent into a magnetic energy-loss spectrometer where it is dispersed, via the Lorentz force, into its spectral components, producing an electron energy-loss (EEL) spectrum. An energy-selecting slit is then used to define an energy window of the spectrum, so that only electrons transmitted through that window are used to form the final image. By moving the energy window across the spectrum a series of energy-loss images can be acquired, leading to a 3D ($x$, $y$, $\Delta E$) data cube [8,18].

EFTEM measurements were carried out at the StEM facility, at the Max Planck Institute for Intelligent Systems in Stuttgart, Germany, using the Zeiss SESAM FEG-TEM [19] operated at 200 kV, fitted with a monochromator and the high-transmissivity in-column MANDOLINE filter [27]. The monochromator was used to achieve an energy spread in the beam of 0.06 eV (FWHM of zero-loss peak). The energy selecting slit used was 0.23 eV wide. EFTEM series were acquired using a routine, written in-house at StEM, using Gatan Digital Micrograph scripts, that allows the automated acquisition of EFTEM series, given the initial and final energy loss and the width of the energy window, adjusting exposure time. Specimen drift was corrected using a Digital Micrograph script described in [28]. The initial and final energies chosen for the measurements presented in this paper are 0.0 eV and 4.0 eV, to cover all the possible low energy excitations between pure elastic scattering (zero energy loss) and the 3.8 eV silver volume plasmon and interband transition [29].

## 3. Results and Discussion.

In this article we report on measurements and analysis of SP resonant modes of a single silver nanorod of aspect ratio 14.2 (diameter 47 nm and length 666 nm), shown in Figure 1(a). The particular geometry of elongated rod particles has been chosen, amongst others present in the nanoparticle solution, for the clarity of the spectral and spatial features excited, as can be seen in the selected-area energy-loss spectra of Figure 1(b) and the acquired EFTEM series in Figure 2. Figure 1(b) shows unprocessed selected area electron energy-loss spectra of different regions along the nanorod, highlighted in Figure 1(a). The spectral features vary dramatically with position along the nanorod with their spatial extent revealed in Figure 2. Note that the spectra at the left and right ends of the nanorod (in blue and red, respectively) are slightly different. This is due primarily to the slight non-isochromaticity across the field of view and seen most clearly in the lowest energy loss image (0.9 eV) in Figure 2 (top left image) where an intensity ramp produces an asymmetry of the image.

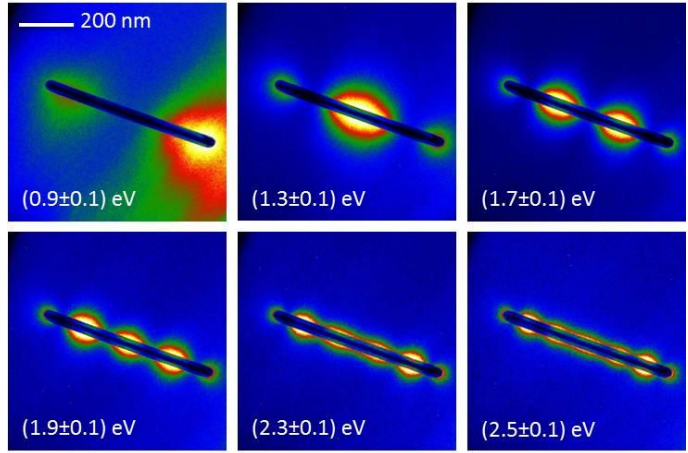

Figure 2. EFTEM series of a single silver nanorod (seen in Figure 1) showing a clear standing-wave pattern due to surface plasmon resonance. The energy selecting slit was 0.23 eV. The images shown here are a subset of the total series and show clear modes from the fundamental ($m=1$ mode) to the $m=6$ mode. The intensity of the energy-loss maps is shown as a temperature colour scale. Due to the strong variation in maximum energy-loss intensity and for better visibility the colour scale has been scaled independently for all 6 images.

The low-loss features seen in Figure 2 have the clear configuration of standing waves, or resonant (cavity) modes. Using a simplistic model we can define the mode index $m$ as being $\lambda_{sp}=2L/m$, where $\lambda_{sp}$ is the wavelength measured on the nanorod and $L$ is the length of the nanorod [1,22-24]. In Figure 2, SP resonant modes from $m=1$ (the 'fundamental' mode) to $m=6$ are clearly discernible in the EFTEM series. Modes with higher $m$ index (i.e. higher energies) are not distinguishable in this series, due to the width of the energy window (0.23 eV) being too large to enable higher frequency modes ($m\geq7$) to be imaged clearly.

The high spatial resolution available with EFTEM maps enables key physical information to be extracted regarding the properties of SP waves at the nanoscale. By measuring the spatial wavelength $\lambda_{sp}$ seen at each energy loss (i.e. for each SP mode) the SP dispersion relation (energy $E$ of the modes as a function of the modulus of the wavevector $\boldsymbol{k}$) can be determined experimentally for an individual nanorod. In the particular case we study, the wavevector is always associated with the waveguide mode of the nanorod guided along the axis of the rod. Thus, we can safely treat the wavevector $\boldsymbol{k}$ as a scalar quantity, the wavenumber $k$. The dispersion relation extracted from the experimental data of Figure 2 is shown in Figures 3(a) and 3(b) (green circles and red crosses). The wavenumber $k$ is determined directly by measuring the distance between two antinodes (maxima of intensity, equal to $\lambda_{sp}/2$) just above and below the nanorod and taking the mean value for a given mode; the wavenumber $k$ is given by $k=2\pi/\lambda_{sp}$. The real and imaginary parts of the wavenumber $k$ are determined by fitting the intensity line profiles of a given mode, assuming that the SP standing-wave can be described by a damped Fabry-Pérot system. The imaginary part of $k$ is determined to be equal to the damping constant that causes standing waves further away from the end points to have smaller intensities.

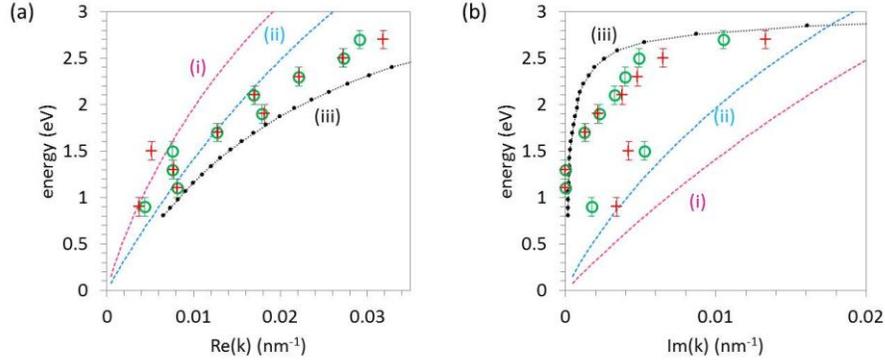

Figure 3. Experimental and calculated dispersion curves for a silver nanorod of length $L=666\pm3$ nm and diameter $d=47\pm3$ nm. (a) Energy loss versus the real part of the wavenumber $k$ extracted from experimental data of Figure 2 (green circles refer to values from the top part of each EFTEM image of the nanorod, red crosses refer to values from the bottom part of each EFTEM image the nanorod) and (i) compared with simulation of an infinitely long silver nanorod (50 nm in diameter) in vacuum, neglecting retardation (dashed pink line), (ii) in vacuum but now considering the effect of retardation (dashed blue line) and (iii) considering both the effect of retardation and the presence of the substrate on which the nanorod is resting (black dots). (b) Energy loss versus the imaginary part of the modulus of the wavenumber $k$. The colour scheme used is the same as in (a).

In order to better understand the underlying physical processes occurring at the nanoscale, we have modelled theoretically the SP dispersion using full electrodynamics simulations obtained by the use of COMSOL Multiphysics, following the description found in [30]. To model the optical behaviour of the sample (constituted by a finite 666 nm long and 47 nm wide silver nanorod resting on 30 nm silicon nitride membrane) we have used, as an initial approximation, an infinitely long, 50 nm wide silver nanorod in vacuum, introducing, as a second step, the effect of retardation [31] and, as a final step to a more realistic model, the presence of the 30 nm thick silicon nitride substrate.

Figures 3(a) and 3(b) show the simulated dispersion curves for the real and imaginary part of the wavenumber $k$. Simulations have been carried out for an infinitely long nanorod of 50 nm diameter in vacuum, taking into consideration retardation and substrate effects (dashed lines (i), (ii) and (iii)). Experimental data extracted from Figure 2 are also shown in Figure 3 as green circles (for values extracted from the top part of each EFTEM image of the nanorod) and red circles (for values extracted from the bottom part of each EFTEM image of the nanorod). As can be seen in Figure 3(a) and (b) the simulations best match experimental data only after both retardation and substrate effects are included, underlining the need to consider the resonance of the whole system, and not just of the nanorod as an independent isolated object.

This is the first result, to our knowledge, where the influence of retardation and substrate effects has been clearly demonstrated by both experiments and simulations for a SP dispersion for an individual nanorod. It is clear that both effects must be considered alongside others already identified in the literature, such as shape, size and composition to understand fully SP dispersion.

## 3.1 Direct measurement of the SP decay length in vacuum.

SP resonances give rise to an oscillating electromagnetic field that is exponentially decaying perpendicular to the relevant metal-dielectric interface [2,5]. The distance at which the intensity of the electric field decreases to a value of 1/e of its initial value is commonly called the "skin depth" inside the metal, and the "decay length" in the dielectric/vacuum surrounding the metal [2]. Knowledge of the decay length is of great importance in certain SP applications, such as in the design of substrates for SERS measurements that take advantage of the enhancement of the electric field between two neighbouring nanoparticles [32,33].

The SP decay length $D$ has been measured from the EFTEM data by plotting the intensity of the energy-loss signal as a function of perpendicular distance from the nanorod axis, for all the maxima visible in the modes discernible in the EFTEM series (Figure 2). The values obtained for the decay length are plotted in Figure 4 as a function of the energy loss.

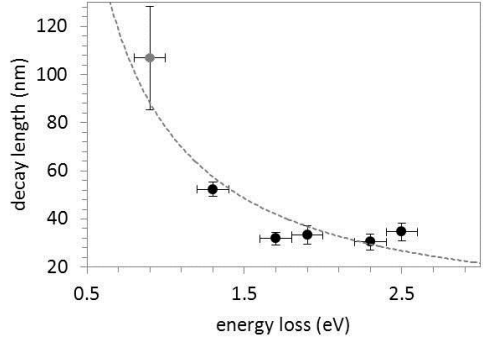

Figure 4. Decay length $D$ as a function of energy loss for the SP resonant modes probed experimentally in Figure 2. Horizontal error bars are defined by the 0.23 eV energy filter window. The dashed grey line fits a power law function with exponent -1.

The value of the decay length at 0.9 eV is based on a mean value measured for the 0.9 eV image. However this is only an approximate determination as this image is particularly prone to the effect of non-isochromaticity which although small (0.06 eV) leads to an intensity ramp across this image as seen in Figure 2. As such, the value is shown in grey. The dashed grey line is a guide to the eye fitted taking into account all the measured decay lengths (including the approximate value obtained for the 0.9 eV measurement) and in accordance with the $1/E$ trend predicted by Egerton for very low loss spectral features [34]. If the 0.9 eV measurement is excluded we find the average decay length $D$ to be 33±8 nm in agreement with values reported previously [2]. Further studies on the energy dependency of the decay length are being undertaken.

### 3.2 The decay of the nanorod tip excitation.

The study of how SP waves behave at discontinuities (in this case the ends of the nanorod) is especially important, as discontinuities can act as emissions centres of electromagnetic radiation in nano-antennae. The EFTEM data seen in Figure 2 reveals an intriguing behaviour of the tip excitation as a function of SP mode. We define the intensity of the excitation as the area of a Gaussian profile fitted to the energy-loss peak of a particular mode in an electron energy-loss spectrum, as seen in Figure 1(b); the Gaussian fitting procedure was performed after removal of the background (modelled as a power law function). Both fitting procedures are part of the Gatan Digital Micrograph EELS Development package. The selected-area spectra have been acquired from the areas shown in Figure 1(a) in blue and red. The diameter of the selected-area circles is 105±3 nm. The intensity measured from equivalent areas of the two ends of the nanorod, spectra in blue and red of Figure 1(b) respectively, for the different modes in Figure 2, follows an exponential decay as plotted in Figure 5. At present it is not known why the exponent should equal -2 but work is on-going to see if this is a universal number across silver nanorods of different lengths and aspect ratio.

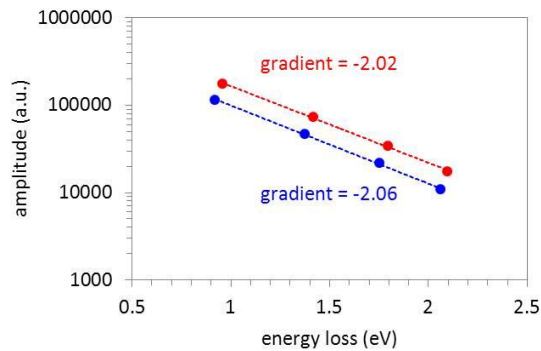

Figure 5. Plot of the intensity (area of the fitted Gaussian of the individual SP peaks in the spectrum of Figure 1(b) in blue and red) at the right tip (red circles) and left tip (blue circles) of the nanorod as a function of energy loss. It is clear that the intensity follows an exponential decay, with exponent of approximately -2. The error in the measurement of intensity is not shown as it is smaller than the size of the markers.

### 3.3 Observations of $\lambda_{sp}$-compression.

As emphasized throughout this paper, EELS based techniques can access spatial details of SP resonant modes at the nanoscale. Taking a close look at the SP standing wave features that appear at different energies in the EFTEM series reported in Figure 2, we observe that, in contrast with the classical case of a standing wave in one dimension, the wavelength, for a given mode, is not constant throughout the length of the nanorod. This is illustrated in Figure 6(a), where the $m$=6 mode of Figure 2 is annotated with the measurements of the antinode spacing, i.e. the half-wavelength $\lambda_{sp}/2$. For all observed modes in the EFTEM series the spatial wavelength $\lambda_{sp}$ has a 'relaxed' value in the central region of the nanorod and is compressed to lower values as the SP wave approaches the end of the nanorod.

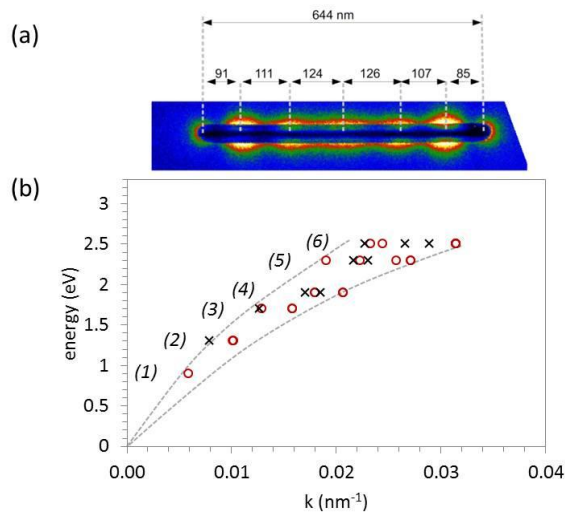

Figure 6. (a) EFTEM map of mode $m$=6 with energy loss window 2.5±0.1 eV. The antinode spacing (half-wavelength $\lambda_{sp}/2$) between adjacent maxima of intensity (antinodes) of the SP standing wave varies along the nanorod. The spacing shown at the top of Figure 5(a) decrease as the wave approaches the ends of the nanorod. (b) Dispersion relation (energy $E$ (in eV) as a function of wavenumber $k$ (in nm$^{-1}$), $k=2\pi/\lambda_{sp}$) measured from the experimental EFTEM series of Figure 2 showing that for a given energy, i.e. for a given resonant mode, the wavenumber $k$ does not have a unique value. Data points indicated in red correspond to the spacing between antinodes (maxima of intensity), whereas data points indicated in black crosses correspond to nodes (minima of intensity). The numbers in brackets indicate the mode number $m$. The dashed grey lines are a guide to the eye.

This effect of $\lambda_{sp}$-compression, or "antinode bunching", has also been observed by Rossouw *et al.* [15] in scanning TEM – EELS (STEM-EELS) experiments on silver nanowires on ultrathin carbon films. An overview of this "$\lambda_{sp}$-compression" effect, observable in our EFTEM experiment, for all modes, is reported in Figure 6(b), where the energy $E$ is plotted as a function of the wavenumber $k$ measured from the data of Figure2. The graph in Figure 6(b) shows how, for a given energy, the wavenumber $k$ has a range of values, highlighting the variation of the wavelength $\lambda_{sp}$ throughout the length of the nanorod. The value of $k$ varies from the lowest, "relaxed", value (i.e. the highest half-wavelength $\lambda_{sp}/2$ distance between maxima in intensity of the energy loss signal) close to the centre of the nanorod, to higher values of $k$ moving away from the centre, towards the ends of the nanorod. Notice also that the antinode with the maximum signal corresponds to the one nearest to, but not at, the end of the nanorod.

The plot in Figure 6(b) shows the values of $k$ subdivided between those measured as the distance between two adjacent antinodes as red circles and those measured as the distance between two adjacent nodes as black crosses. Note the presence of an antinode excitation at the end of the nanorod throughout the energy loss range.

We speculate that this effect of $\lambda_{sp}$-compression is induced by the reflection mechanism of the SP wave at the ends of the nanoparticle, being dependent on the size and shape of the ends, on the surrounding medium outside the ends and, at a more atomistic level, on the surface-electron scattering. BEM simulations on finite silver nanorods, that take into account these parameters, and the building of a full theoretical model of such an effect, are currently being undertaken and will be the subject of a future publication.

### 4. Conclusions.

Direct mapping of surface plasmon standing waves on a silver nanorod has been achieved using energy-filtered TEM. The use of a parallel monochromated electron beam has enabled the analysis of SP resonance with high spatial, and high energy, resolution. EFTEM maps have been used to observe and measure important and novel physical properties of SP resonant modes of metal nanoparticles. Comparison between experimental data and full electrodynamics simulations of a silver nanorod sitting on a silicon nitride membrane has revealed how important it is to include retardation and substrate effects in understanding fully the nature of a given SP resonant mode of the nanorod. The measurement of the decay length of a SP, the energy-dependent amplitude excitation at the ends and the analysis of $\lambda_{sp}$-compression show how EELS-based techniques (EFTEM in this case) can reveal fine-scale spatial features that are difficult to determine with other methods and can lead to a greater understanding of the nanoscale physics underlying SP resonant modes of metallic nanoparticles.

**Acknowledgments**
ON, WS, PvA and PAM acknowledge financial support from the European Union under the Framework 6 program under a contract for an Integrated Infrastructure Initiative. Reference 026019 ESTEEM. MW acknowledges financial support from the Danish Research Council for Technology and Production Sciences (FTP grant #274 – 07 – 0080).